%% file: main.tex
\documentclass[sigconf,natbib=true]{acmart}
\AtBeginDocument{%
  \providecommand\BibTeX{{%
    \normalfont B\kern-0.5em{\scshape i\kern-0.25em b}\kern-0.8em\TeX}}}

\usepackage{booktabs} % To thicken table lines

\usepackage[english]{babel}
\usepackage{moresize}
\usepackage{threeparttable} 
\usepackage{amsmath}
\usepackage{algorithmic}
\usepackage{balance}
\usepackage{comment}
\usepackage{paralist}
\usepackage{bm}
\usepackage{bbding}
\usepackage{pgfplots}
\usetikzlibrary{pgfplots.dateplot}
\usepackage{xcolor}
\usepackage{flushend}
\usepackage[english]{babel}
\usepackage[latin1]{inputenc}
\usepackage{mathrsfs}
\usepackage{graphicx}

\usepackage{amssymb}
\usepackage{amsfonts}
\usepackage{url}
\usepackage{longtable}
\usepackage{rotating}
\usepackage{multirow}
\usepackage{mathrsfs}
\usepackage{subfigure}
\usepackage{enumitem}
\usepackage[linesnumbered,algoruled,boxed,lined]{algorithm2e}
\usepackage{adjustbox}
\usepackage{hyperref}
\usepackage{pgfplots}
\usetikzlibrary{pgfplots.dateplot}
\usepackage{filecontents}
\usepackage{tabularx}
\usepackage{threeparttable}
\definecolor{tblue}{RGB}{31,119,180}
\definecolor{torange}{RGB}{255,127,14}
\definecolor{tgreen}{RGB}{44,160,44}
\definecolor{tred}{RGB}{214,39,40}
\definecolor{tpurple}{RGB}{148,103,189}

\usepackage{colortbl}
\usepackage{xcolor}
\usepackage{array}

\newcommand{\hide}[1]{} %hide

\newcommand{\ie}{\textit{i}.\textit{e}.}
\newcommand{\eg}{\textit{e}.\textit{g}.}

\def\model{SSLRec}

\setcopyright{none}

\copyrightyear{2024}
\acmYear{2024}
\setcopyright{acmlicensed}
\acmConference[WSDM '24] {Proceedings of the 17th ACM International Conference on Web Search and Data Mining}{March 4--8, 2024}{Merida, Mexico.}
\acmBooktitle{Proceedings of the 17th ACM International Conference on Web Search and Data Mining (WSDM '24), March 4--8, 2024, Merida, Mexico}
\acmPrice{15.00}
\acmISBN{979-8-4007-0371-3/24/03}
\acmDOI{10.1145/3616855.3635814}

\begin{document}

\title{SSLRec: A Self-Supervised Learning Framework \\ for Recommendation}

\author{Xubin Ren}
\affiliation{%
  \institution{University of Hong Kong}
  \city{}
  \country{}}
\email{xubinrencs@gmail.com}

\author{Lianghao Xia}
\affiliation{%
  \institution{University of Hong Kong}
  \city{}
  \country{}}
\email{aka_xia@foxmail.com}

\author{Yuhao Yang}
\affiliation{%
  \institution{University of Hong Kong}
  \city{}
  \country{}}
\email{yuhao-yang@outlook.com}

\author{Wei Wei}
\affiliation{%
  \institution{University of Hong Kong}
  \city{}
  \country{}}
\email{weiweics@connect.hku.hk}

\author{Tianle Wang}
\affiliation{%
  \institution{University of Hong Kong}
  \city{}
  \country{}}
\email{louiswong.cs@connect.hku.hk}

\author{Xuheng Cai}
\affiliation{%
  \institution{University of Hong Kong}
  \city{}
  \country{}}
\email{rickcai@connect.hku.hk}

\author{Chao Huang}
\authornote{Corresponding author.}
\affiliation{%
  \institution{University of Hong Kong}
  \city{}
  \country{}}
\email{chaohuang75@gmail.com}

\renewcommand{\shorttitle}{SSLRec: A Self-Supervised Learning Framework for Recommendation}
\renewcommand{\shortauthors}{Xubin Ren et al.}

\begin{abstract}

Self-supervised learning (SSL) has gained significant interest in recent years as a solution to address the challenges posed by sparse and noisy data in recommender systems. Despite the growing number of SSL algorithms designed to provide state-of-the-art performance in various recommendation scenarios (\eg, graph collaborative filtering, sequential recommendation, social recommendation, KG-enhanced recommendation), there is still a lack of unified frameworks that integrate recommendation algorithms across different domains. Such a framework could serve as the cornerstone for self-supervised recommendation algorithms, unifying the validation of existing methods and driving the design of new ones. To address this gap, we introduce \model, a novel benchmark platform that provides a standardized, flexible, and comprehensive framework for evaluating various SSL-enhanced recommenders. The \model\ framework features a modular architecture that allows users to easily evaluate state-of-the-art models and a complete set of data augmentation and self-supervised toolkits to help create SSL recommendation models with specific needs. Furthermore, \model\ simplifies the process of training and evaluating different recommendation models with consistent and fair settings. Our \model\ platform covers a comprehensive set of state-of-the-art SSL-enhanced recommendation models across different scenarios, enabling researchers to evaluate these cutting-edge models and drive further innovation in the field. Our implemented \model\ framework is available at the 
source code repository \color{blue}\url{https://github.com/HKUDS/SSLRec}.
\end{abstract}



\maketitle
\input{intro}
\input{sslrecsys}
\input{framework}

\input{benchmark}
\input{relatedwork}
\input{conclusion}

\clearpage
\balance
\bibliographystyle{ACM-Reference-Format}
\bibliography{sample-base}

% \appendix

\end{document}

%% file: intro.tex
\section{Introduction}
\label{sec:intro}

Recommender systems have become indispensable in tackling the challenge of information overload and enhancing user experiences across a wide range of online personalized services~\cite{zhang2020explainable,wu2022graph,rendle2020neural}. In recent years, deep learning techniques have achieved remarkable success, and various neural network techniques have been widely adopted to enhance the performance of recommender systems~\cite{zhang2019deep}. However, the effectiveness of many existing methods can be limited by the availability of high-quality supervision labels, particularly in recommender systems where large numbers of items can result in sparse user interactions and insufficient training labels for many long-tail users and items~\cite{liu2020long,sankar2021protocf}. Additionally, noise in user behavior data, such as interaction noise~\cite{tian2022learning} and popularity bias~\cite{chen2023bias}, is a pervasive challenge that significantly affects the model robustness, leading to a degradation in the recommendation performance.

As the need for recommender systems that can learn from sparse and noisy supervision signals grows, self-supervised learning (SSL) approaches have emerged and shown promising results in reducing dependence on observed supervision labels~\cite{liu2021self,yu2022self}. However, many SSL recommender systems lack public implementations, and even with open-source code, inconsistencies in implementation details such as data format, parameter tuning methods, and model training strategies across different developers and settings make it challenging to compare the effectiveness of various algorithms fairly. This inconsistency impedes progress in the field of recommendation. To address this challenge, we introduce the \model\ framework, which offers a standardized, flexible, and comprehensive platform for comparing the performance of different SSL-enhanced recommender systems. Our framework provides a unified approach that covers various recommendation scenarios, enabling researchers and practitioners to evaluate the effectiveness of different algorithms consistently and fairly. By identifying the strengths and weaknesses of different algorithms, our \model\ framework provides valuable insights for further research and development, thus facilitating the advancement of the field of recommender systems. We highlight the key features of our framework from the following perspectives:

\begin{itemize}[leftmargin=*]\vspace{-0.05in}

\item \textbf{Flexible Modular Architecture:} Our \model\ boasts a modular architecture that enables users to tailor and merge various modules with ease, thereby creating their own customized recommendation models. Moreover, the framework offers an extensive selection of pre-built modules that can be seamlessly integrated into different models across various scenarios. With this modular architecture, users can enjoy greater flexibility in designing and testing models, catering to their specific needs and preferences.

\item \textbf{Diverse Recommendation Scenarios}. Our \model\ framework is designed to cater to a wide range of recommendation scenarios, making it a valuable tool for researchers and practitioners seeking to construct effective recommendation models in various domains of recommender systems. Our framework comprehensively addresses various aspects of user preferences, such as graph-based collaboration, sequential item transitions, social-awareness, heterogeneous interactions, and knowledge-awareness. This broad coverage allows users to apply our framework in diverse recommendation scenarios, enabling the development of superior recommendation models and supporting various applications and research lines.

\item \textbf{Comprehensive State-of-the-Art Models}. The \model\ framework is a comprehensive collection of state-of-the-art SSL-enhanced recommendation models that are designed for various scenarios. This framework has several advantages, including its ability to enable researchers to evaluate these cutting-edge models using advanced techniques that are at the forefront of SSL investigation for recommendation. Additionally, the \model\ framework is an essential tool for advancing the state-of-the-art in SSL-enhanced recommendation, as it provides a comprehensive and easily accessible platform for researchers to develop new models.

\item \textbf{Unified Data Feeder and Standard Evaluation Protocols}. The \model\ framework features a unified data feeder and standard evaluation protocols, making it easy for users to load and preprocess data from different sources and formats. With a standardized evaluation process, the \model\ framework provides a consistent, objective, and fair way to evaluate the performance of various recommendation models. The \model\ framework's design ensures that users can evaluate and compare SSL-enhanced recommender systems in a reliable and efficient manner.

\item \textbf{Rich Utility Functions}. The \model\ framework offers a wide range of utility functions that simplify the development and evaluation of recommendation models. We have implemented commonly used functionalities of recommender systems and self-supervised learning, such as graph operations, neural network architectures, and loss functions, to help researchers integrate new models and improve performance evaluation. By leveraging these utility functions, researchers can focus on developing innovative algorithms instead of spending time implementing basic functionalities from scratch. It streamlines development, reduces time, and optimizes resource utilization.

\item \textbf{Easy-to-Use Interface}: The \model\ framework offers an easy-to-use interface that simplifies the process of training and evaluating recommendation models. Our user-friendly interface enables researchers and practitioners to experiment with different models and configurations quickly and easily. The \model\ framework's easy-to-use interface makes it accessible to a wide range of users and facilitates fast and efficient experimentation with different SSL-enhanced recommender systems.

\end{itemize}

Although open-source recommender system frameworks such as ReChorus~\cite{wang2020make}, RecBole~\cite{zhao2021recbole}, and Beta-rec~\cite{meng2020beta} have recently been developed, the \model\ stands out by focusing specifically on self-supervised learning (SSL)-enhanced recommender systems, which represent the current state-of-the-art in recommendation research. Additionally, unlike the most recent framework for SSL-based recommenders~\cite{yu2022self}, which primarily focuses on the static scenario of graph collaborative filtering, our \model\ targets a more diverse range of recommendation scenarios and provides a comprehensive set of SSL-enhanced recommendation models. Our framework features a rich set of utility functions to facilitate framework usage. The \model\ framework's unique and specialized focus on SSL-enhanced recommender systems makes it an essential tool for researchers and practitioners who want to stay up-to-date with the latest advancements in the field of recommender systems.

%% file: sslrecsys.tex
\section{Self-supervised Recommenders}
\label{sec:sslrecsys}

We will start by introducing the basics of recommender systems in various scenarios and then explain how they can be enhanced by self-supervised learning. Generally, recommender systems consist of a set of $M$ users denoted by $\mathcal{U}={u}$ and a set of $N$ items denoted by $\mathcal{I}={i}$. In different recommendation scenarios, the observation data $\mathcal{X}$ generated from users and items varies.

\begin{itemize}[leftmargin=*]

\item \textbf{General Collaborative Filtering (CF)} is a widely adopted paradigm in personalized recommendation systems that takes into account the similarity between users based on their interaction behaviors. This is accomplished by mapping users and items into a low-dimensional latent embedding space, leveraging their past interactions. Recent studies~\cite{he2020lightgcn,wu2021self,wang2022towards,jiang2023adaptive} employ Graph Neural Network (GNN)-based message passing techniques to capture intricate user-item interaction patterns. The observed data $\mathcal{X}$ consists of user-item interactions represented as a bipartite graph, with each edge indicating a user-item interaction. \\\vspace{-0.12in}

\item \textbf{Multi-Behavior Recommendation} aims to harness the power of multiple user behaviors, including clicks, add-to-favorite, reviews, and purchases, to produce precise recommendations. By treating each behavior type individually, multi-behavior recommenders can effectively capture the nuanced preferences associated with each behavior. This enables a more holistic comprehension of user-item interactions, taking into account the distinctive preferences expressed through different types of user behaviors. As a result, the user-item interaction data $\mathcal{X}$ encompasses a rich tapestry of heterogeneous relationships between users and items. \\\vspace{-0.12in}

\item \textbf{Sequential Recommendation} involves considering the temporal and sequential aspects of user behavior when generating recommendations~\cite{sun2019bert4rec,xie2022contrastive,qiu2022contrastive}. In this context, the observation data $\mathcal{X}$ comprises a collection of interaction sequences ordered chronologically, incorporating both the interaction behaviors and their temporal information. The primary objective is to predict the next item in each individual user's interaction sequence, which requires a deep understanding of user preferences and the underlying patterns in their temporal behavior. \\\vspace{-0.12in}

\item \textbf{Social Recommendation} is a research line that leverages user-user relationships, such as friendships or followerships, as additional social information to enhance the understanding of user preferences~\cite{huang2021knowledge,du2022socially,chen2023heterogeneous}. By incorporating these social connections into the recommendation process, social recommendation systems aim to improve the accuracy and relevance of recommendations. In social recommendation, the observation data $\mathcal{X}$ encompasses not only the interactions between users and items but also a social graph that captures the relationships between users. \\\vspace{-0.12in}

\item \textbf{Knowledge Graph-enhanced Recommendation} involves integrating knowledge graphs into the recommender systems to leverage the wealth of semantic information associated with items~\cite{huang2021knowledge, yang2023knowledge}. In this line, the observation data $\mathcal{X}$ consists of two interconnected graphs: the user-item interaction graph and an auxiliary knowledge graph. The knowledge graph utilizes entity-relation-entity triplets to categorize item attributes based on the types of entities involved and their relationships.

\end{itemize}

In general, a recommendation model equipped with the predictive function $f(\cdot)$ aims to estimate the likelihood $f(\mathcal{X})_{u,i}$ of a user $u$ engaging with an item $i$. This estimation process is formulated as an optimization problem, which takes into account the ground truth of user-item interactions $\mathcal{Y}$, where each interaction $y_{u,i}\in\mathcal{Y}$ signifies a known user-item pair. Additionally, a recommendation loss function $\mathcal{L}_\text{rec}$, such as BPR loss~\cite{rendle2012bpr}, hinge loss~\cite{xia2021knowledge}, or cross-entropy loss~\cite{he2016fast, he2017neural}, is employed to measure the dissimilarity between the predicted likelihood and the ground truth.
\begin{align}
    \label{eq:rec}
    \min \sum_{y_{u,i}\in\mathcal{Y}} \mathcal{L}_\text{rec}(f(\mathcal{X})_{u, i}, y_{u,i})
\end{align}

To enhance the representation capabilities of recommendation models when dealing with sparse data, self-supervised learning (SSL) techniques have gained significant attention in all the aforementioned scenarios. The fundamental concept involves incorporating auxiliary self-supervised tasks alongside the traditional supervised task of recommendation to augment the available data. Three commonly employed SSL paradigms in recommendation are contrastive, generative, and predictive self-supervision.

\begin{itemize}[leftmargin=*]
    \item \textbf{Contrastive SSL} contrasts embeddings encoded from different augmentation views is a key component of self-supervised learning (SSL) for recommendation~\cite{wu2021self,xia2022hypergraph}. It aims to bring similar examples closer together in the representation space while pushing dissimilar examples farther apart. Various augmentation view generation schemes are employed for embedding contrasting: i) Stochastic structure corruption is applied to user interactions (\eg, SGL~\cite{wu2021self}), sequences (\eg, CL4Rec~\cite{xie2022contrastive}), and knowledge graphs (\eg, KGCL~\cite{yang2022knowledge}). Local-global contrastive augmentation incorporates global information using methods like HCCF~\cite{xia2022hypergraph} and NCL~\cite{lin2022improving}. Contrastive augmentation across diverse user-item interactions is achieved with techniques such as CML~\cite{wei2022contrastive}.

    \item \textbf{Generative SSL} employ a generative model and a reconstruction objective for self-supervision. They aim to reconstruct input data in various scenarios: i) In sequence encoding, models like BERT4Rec~\cite{sun2019bert4rec} and MAERec~\cite{ye2023graph} reconstruct masked items within sequences and item graphs, respectively. For the user-item interaction graph, approaches such as AutoCF~\cite{xia2023automated} utilize masked autoencoders to reconstruct the data.

    \item \textbf{Predictive SSL} utilize specific properties or attributes of the data as self-generated labels, as self-supervised signals, for prediction. Examples include: i) In social recommendation, methods like SMIN~\cite{long2021social} and MHCN~\cite{yu2021self} introduce a node-graph relation predictor to capture the global graph context. This predictor leverages self-supervised signals to enhance prediction tasks in social recommendation scenarios. For collaborative filtering tasks, techniques such as SHT~\cite{sht2022} incorporate an Interaction denoising module, to improve the collaborative filtering task through predictive self-supervision.
    
\end{itemize}

The SSL paradigm encompasses various techniques, including contrastive SSL, generative SSL, and predictive SSL, which all share the characteristic of utilizing augmented data views. These views can take the form of two embedding views in contrastive SSL, masked and original data views in generative SSL, or data views used for estimating mutual information in predictive SSL. Based on this common observation, we can summarize the SSL paradigm as an approach that leverages augmented data views to enhance the learning process with respect to the data observation $\mathcal{X}$.
\begin{align}
    \label{eq:ssl}
    \min \mathcal{L}_\text{ssl}(\Delta_1(\mathcal{X}), \Delta_2(\mathcal{X}))
\end{align}
\noindent $\Delta_1(\cdot)$ and $\Delta_2(\cdot)$ refer to specific \textbf{data augmentation methods} used for generating the corresponding data views. The \textbf{SSL objective}, $\mathcal{L}_\text{ssl}$, optimizes the model based on the augmented views.

Our proposed framework, \model, draws on insights from various recommender systems and self-supervised learning approaches to provide a cohesive and modular solution for implementing SSL methods in different scenarios. This framework offers a high degree of modularity in data augmentation and general SSL functions, making it efficient for implementing SSL approaches based on the general paradigm defined by Eq~\ref{eq:rec} and Eq~\ref{eq:ssl}. Furthermore, it offers a unified training and evaluation process that ensures both reproducibility and fair comparison of results.

%% file: framework.tex
\section{The SSLRec Framework}
\label{sec:framework}

In this section, we introduce the \model, which is built on PyTorch~\cite{paszke2019pytorch}. Our framework provides a standard and unified design for implementing SSL methods in recommendation. To date, the SSLRec framework has incorporated around 30 state-of-the-art self-supervised methods across five distinct recommendation scenarios. Additionally, the framework offers various SSL-related modules to support the development of recommender systems.

\begin{table}
    \centering
    \small
    %\footnotesize
    %\scriptsize
    \caption{SOTA SSL-based Recommenders Implemented in \model\ across five diverse recommendation scenarios.}
    \vspace{-0.15in}
    \begin{tabular}{ccccc}
        \toprule
        Scenarios & Models & Venue & Year\\
        \midrule
        \multirow{10}{*}{General Collaborative Filtering} 
        & LightGCN~\cite{he2020lightgcn}         & SIGIR & 2020\\
        & SGL~\cite{wu2021self}                  & SIGIR & 2021\\
        & HCCF~\cite{xia2022hypergraph}          & SIGIR & 2022\\
        & SimGCL~\cite{yu2022graph}              & SIGIR & 2022\\
        & NCL~\cite{lin2022improving}            & WWW   & 2022\\
        & DirectAU~\cite{wang2022towards}        & KDD   & 2022\\
        & LightGCL~\cite{cailightgcl}            & ICLR  & 2023\\
        & AutoCF~\cite{xia2023automated}         & WWW   & 2023\\
        & DCCF~\cite{ren2023disentangled}        & SIGIR & 2023\\
        & GFormer~\cite{li2023graph}             & SIGIR & 2023\\
        \midrule
        \multirow{5}{*}{Sequential Recommendation} 
        & BERT4Rec~\cite{sun2019bert4rec}        & CIKM  & 2019\\
        & CL4SRec~\cite{xie2022contrastive}      & ICDE  & 2022\\
        & DuoRec~\cite{qiu2022contrastive}       & WSDM  & 2022\\
        & ICLRec~\cite{chen2022intent}           & WWW   & 2022\\
        & DCRec~\cite{dcrec2023}                 & WWW   & 2023\\
        \midrule
        \multirow{4}{*}{Social Recommendation} 
        & KCGN~\cite{huang2021knowledge}         & AAAI  & 2021\\
        & SMIN~\cite{long2021social}             & CIKM  & 2021\\
        & MHCN~\cite{yu2021self}                 & WWW   & 2021\\
        & DSL~\cite{wang2023denoised}            & IJCAI & 2023\\
        \midrule
        \multirow{3}{*}{\shortstack{Knowledge Graph-enhanced \\ Recommendation} } 
        & KGIN~\cite{wang2021learning}           & WWW   & 2021\\
        & KGCL~\cite{yang2022knowledge}          & SIGIR & 2021\\
        & KGRec~\cite{yang2023knowledge}         & KDD   & 2023\\
        \midrule
        \multirow{5}{*}{Multi-behavior Recommendation} 
        & MBGMN~\cite{xia2021graph}              & SIGIR & 2021\\
        & HMG-CR~\cite{yang2021hyper}            & ICDM  & 2021\\
        & S-MBRec~\cite{gu2022self}              & IJCAI & 2022\\
        & CML~\cite{wei2022contrastive}          & WSDM  & 2022\\
        & KMCLR~\cite{xuan2023knowledge}         & WSDM  & 2023\\
        \bottomrule
    \end{tabular}
    \vspace{-0.25in}
    \label{tab:implemented_models}
\end{table}

\begin{figure*}
    \centering
    \includegraphics[width=1.0\textwidth]{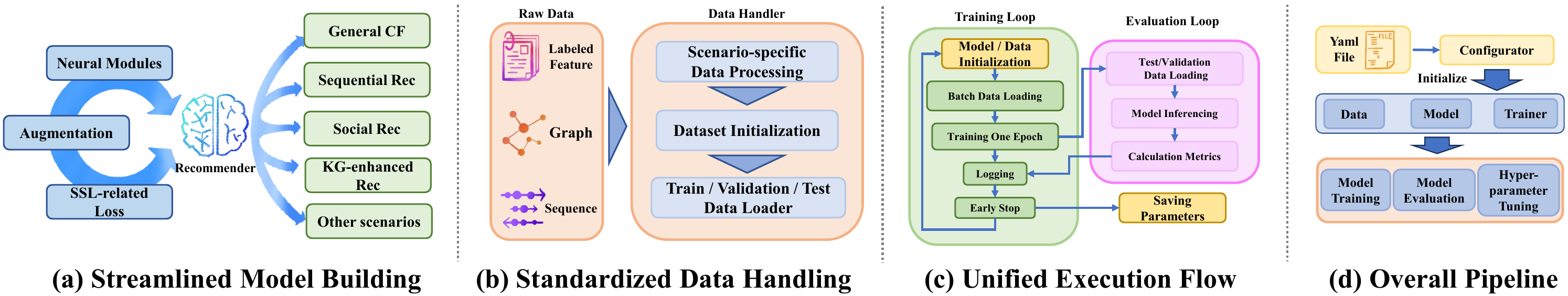}
    \vspace{-0.25in}
    \caption{The overall framework of our proposed \model.}
    \vspace{-0.15in}
    \label{fig:framework}
\end{figure*}

\subsection{Modular and Standardized Models}

A reproducible and sustainable recommender framework relies on the implementation of modular and standardized algorithms. As stated earlier, both augmentation methods and self-supervised objectives are essential components of most self-supervised recommender models across different recommendation scenarios. To address this, we distilled the augmentation patterns from various SSL algorithms and encapsulated them into model-agnostic modules. Additionally, we standardized commonly used loss functions in SSL to enable their use across different algorithms. To date, we have implemented approximately 30 state-of-the-art SSL methods across five scenarios, as shown in Table~\ref{tab:implemented_models}, using our reusable SSL-related modules. \\\vspace{-0.2in}

\subsubsection{\bf Augmentation Modules} To prevent model degradation caused by sparse labels, data augmentation plays a crucial role in generating diverse perspectives of raw data or features, facilitating the creation of supplementary self-supervised signals. In \model, we distill commonly used augmentation techniques across various SSL algorithms and structure them into model-agnostic modules for easy implementation across diverse SSL models. Augmentation techniques for recommender systems can be broadly classified into two categories: \textbf{data-based} and \textbf{feature-based} methods. \\\vspace{-0.05in}

\noindent \textbf{Data-based Augmentation.} Data-based augmentation methods for graph-based recommenders include stochastic augmentation~\cite{wu2021self} of graph structures, such as randomly dropping edges or nodes, and adaptive augmented graph structure learning~\cite{xia2023automated,jiang2023adaptive}. In social recommendation scenarios, researchers utilize matrix operations to obtain an augmented interaction matrix by mining specific interaction patterns~\cite{yu2021self,yu2021socially} between social relations. This reduces noise and enriches the augmented view with social-aware semantic information. For sequential scenarios, heuristic or random enhancements~\cite{xie2022contrastive, qiu2022contrastive}, such as cropping, insertion, and reordering of the original item sequence, can be applied to augment the raw data.

\noindent \textbf{Feature-based Augmentation.} In addition to data-based augmentation methods, feature-based techniques aim to create different views based on learned embeddings. These techniques include adding random noise to features~\cite{yu2022graph}, randomly dropping some elements in hidden variables, or learning cluster center clusters of features~\cite{lin2022improving}. However, existing recommendation algorithm frameworks~\cite{zhao2021recbole, yu2022self} do not fully integrate different SSL-relevant augmentation paradigms. In SSLRec, we incorporate diverse augmentation techniques, including those mentioned above, into reusable modules, facilitating their plug-and-play integration. This enables the rapid reproduction of existing SSL algorithms and inspires researchers to devise novel augmentation approaches to further improve model performance and generalization ability.

\subsubsection{\bf Self-Supervised Learning Objectives}
Self-supervised recommendation algorithms leverage the data and features obtained through augmentation by using specific objective functions to generate self-supervised signals that promote the optimization of model parameters. In \textbf{Contrastive/Predictive SSL}, the InfoNCE~\cite{you2020graph} loss is commonly employed to pull the views of positive pairs closer together while simultaneously pushing away the views of negative pairs. There also exist other types of SSL loss. For example, DirectAU~\cite{wang2022towards} proposed alignment and uniformity regularization techniques to directly optimize the distribution of hidden features instead of using the InfoNCE. Additionally, KL/JS-divergence~\cite{wang2022self} is used as an SSL strategy to bring the probability distributions of model predictions closer together, thus achieving self-supervised learning. In \textbf{Generative SSL}, the Cross Entropy loss is widely used in sequential recommendation to align the predicted distribution of masked items with the ground truth. For reconstructing masked edges or node features in a graph, Scaled Cosine Error (SCE)~\cite{hou2022graphmae, tian2023heterogeneous} or contrastive reconstruction~\cite{ye2023graph} have been proposed as more efficient approaches in SSL graph-based encoders.

Our model framework incorporates a comprehensive selection of SSL-related objectives, including the ones mentioned above, as standard loss functions for model implementation. This integration ensures the effectiveness of self-supervised learning while simultaneously improving the overall efficiency of model implementation.

\subsubsection{\bf Standardized Interface} To ensure a cohesive design for the entire framework, we have standardized the public interfaces of all implementation models. This standardization greatly simplifies the process of calling model modules from other modules within the framework, such as the trainer and evaluator. By utilizing the SSL sub-modules and functions mentioned earlier, each algorithm is mandated to include the following trio of public interfaces:
\vspace{-0.05in}

\begin{itemize}[leftmargin=*]
\item \textbf{\textsf{forward}$(\cdot)$}. It represents the model-specific forward process (\eg, message passing in graph-based methods) that takes raw features as input and generates the required embeddings for prediction.

\item \textbf{\textsf{cal\_loss}$(\cdot)$}. This function is for making predictions based on the embeddings generated by the \textsf{forward}$(\cdot)$ function and calculating the corresponding loss function values for optimization.

\item \textbf{\textsf{full\_predict}$(\cdot)$}. This function is utilized for evaluation purposes, where the model predicts preference scores for all items based on the final embeddings generated by the \textsf{forward}$(\cdot)$ function.
\vspace{-0.1in}
\end{itemize}

\subsection{Unified Algorithm Execution Process}

\model\ creates a level playing field for training and validating advanced self-supervised learning recommendation algorithms, while aiding researchers in developing novel models. This is achieved by consolidating model-agnostic data processing, training verification, and hyper-parameter tuning into unified execution processes, enabling efficient and fair utilization of different algorithms.

\begin{table}
    \centering
    \small
    %\footnotesize
    %\scriptsize
    \caption{Statistics of experimental datasets in \model.}
    \vspace{-0.15in}
    \begin{tabular}{ccc}
        \toprule
        Scenarios & Data & Size of Observation Data $\mathcal{X}$\\
        \midrule
        \multirow{6}{*}{\shortstack{General CF}} 
        & \multirow{2}{*}{Yelp} & \multirow{2}{*}{\shortstack{\# users: 42,712, \# items: 26,822,\\\# interactions: 182,357}}\\\\
        \cline{2-3}
        & \multirow{2}{*}{Gowalla} & \multirow{2}{*}{\shortstack{\# users: 25,557, \# items: 19,747,\\\# interactions: 294,983}}\\\\
        \cline{2-3}
        & \multirow{2}{*}{Amazon} & \multirow{2}{*}{\shortstack{\# users: 76,469, \# items: 83,761,\\\# interactions: 966,680}}\\\\
        \midrule
        \multirow{4}{*}{\shortstack{Sequential \\ Recsys}} 
        & \multirow{2}{*}{\shortstack{MovieLens\\\-20M}} & \multirow{2}{*}{\shortstack{\# users: 96,726, \# items: 10,154,\\ \# interactions: 193,452}}\\\\
        \cline{2-3}
        & \multirow{2}{*}{Sports} & \multirow{2}{*}{\shortstack{\# users: 85,226, \# items: 56,974,\\ \# interactions: 170,452}}\\\\
        \midrule
        \multirow{4}{*}{\shortstack{Social \\ Recsys}} 
        & \multirow{2}{*}{\shortstack{Yelp\\(Social)}} & \multirow{2}{*}{\shortstack{\# users: 43,043, \# items: 66,576, \\ \# interactions: 283,512}}\\\\
        \cline{2-3}
        & \multirow{2}{*}{Epinions} & \multirow{2}{*}{\shortstack{\# users: 18,081, \# items: 251,722, \\ \# interactions: 715,821}}\\\\
        \midrule
        \multirow{12}{*}{\shortstack{KG-enhanced \\
Recsys}} 
        & \multirow{4}{*}{\shortstack{Book}} & \multirow{4}{*}{\shortstack{\# users: 70,679, \# items: 24,915, \\ \# interactions: 846,434, \\\# relations: 39,\# entities: 29,714, \\\# triplets: 686,516}}\\\\\\\\
        \cline{2-3}
        & \multirow{4}{*}{Alibaba} & \multirow{4}{*}{\shortstack{\# users: 114,737, \# items: 30,040, \\ \# interactions: 1,781,093, \\\# relations: 51,\# entities: 59,156, \\\# triplets: 279,155}}\\\\\\\\
        \cline{2-3}
        & \multirow{4}{*}{MIND} & \multirow{4}{*}{\shortstack{\# users: 100,000, \# items: 30,577, \\ \# interactions: 2,975,319, \\\# relations: 512,\# entities: 24,733, \\\# triplets: 148,568}}\\\\\\\\
        \midrule
        \multirow{10}{*}{\shortstack{Multi-behavior \\
Recsys}} 
        & \multirow{3}{*}{\shortstack{IJCAI}} & \multirow{3}{*}{\shortstack{\# users: 22,438, \# items: 35,573,\\\# click: 1,420,092, \# cart: 1,330,\\ \# fav: 158,119, \# buy: 199,654}}\\\\\\
        \cline{2-3}
        & \multirow{4}{*}{Tmall} & \multirow{4}{*}{\shortstack{\# users: 31,882, \# items: 31,232,\\\# page-view: 1,093,807, \\\# cart: 140,068,\# fav: 49,482, \\\# buy: 167,862}}\\\\\\\\
        \cline{2-3}
        & \multirow{3}{*}{Retail-rocket} & \multirow{3}{*}{\shortstack{\# users: 2,174, \# items: 30,113, \\\# buy: 9,551\# view: 75,374, \\\# cart: 12,456}}\\\\\\
        
        \bottomrule
    \end{tabular}
    \vspace{-0.2in}
    \label{tab:dataset}
\end{table}

\subsubsection{\bf Data Flow}
To ensure reusability, we have established a standardized data flow schema that provides data for training, validation, and testing of recommenders in various scenarios with different features and labels. Table~\ref{tab:dataset} shows representative datasets for the five recommendation scenarios, each featuring different observation data $\mathcal{X}$. We apply standard pre-processing techniques such as removing low-rating interactions and k-core regularization to ensure the quality of each dataset. Our unified data flow uniformly processes data across different recommendation scenarios. The DataHandler class loads the raw data (\eg, user-item interaction matrix) and transforms it into easily usable data structures. The Dataset class creates torch.utils.data.DataLoader objects for training and evaluation purposes, and the resulting features and labels are passed to the model module. The DataHandler class provides a unified external interface for easy integration with other modules. By following this standardized data flow, \model\ ensures consistent and efficient handling of data across all scenarios. \vspace{-0.05in}

\subsubsection{\bf Training and Evaluation}
Our \model\ offers a comprehensive trainer that manages the entire training process, encompassing epoch training, logging, evaluation, testing, metric computation, and model saving. By leveraging the unified public interfaces provided by other modules in \model, our trainer adheres to a standardized training procedure that ensures a fair evaluation and comparison of the effectiveness of various SSL recommenders across all recommendation scenarios. For models with specific training requirements, developers can write essential PyTorch-style code to override specific functions in the trainer. This approach prevents code repetition, maintains consistency, and allows customization while ensuring fairness and coherence within the framework.

\subsubsection{\bf Hyperparameter Tuning} 
In \model, all model configurations, including model hyperparameters, data selection, and trainer configuration, are specified in a yaml file. The robust Configurator module within \model\ initializes the necessary instances based on this yaml file, ensuring a unified and consistent configuration process for all recommendation scenarios. Moreover, we have developed a powerful tuner module that utilizes the unified configuration interface of the yaml file to conduct a depth-first search for optimal hyperparameters via grid search. Researchers only need to define the desired search range in the configuration file, and the tuner module takes care of the rest. This approach simplifies the hyperparameter tuning process and guarantees reproducibility across various experiments, enhancing the efficiency and reliability of model development within the \model\ framework.

%% file: benchmark.tex
\section{Benchmarking}
\label{sec:benchmark}

In this section, we present the experimental results of SSLRec across a range of recommendation tasks. Through our user-friendly interfaces and standardized training and evaluation procedures, we have established a dependable and reproducible leaderboard for each recommendation scenario. This leaderboard effectively tracks the state-of-the-art results achieved by various methods.

\subsection{General Collaborative Filtering}

\begin{table}
    \centering
    \small
    %\footnotesize
    %\scriptsize
    \caption{Model performance reproduced by \model\ for General Collaborative Filtering on Gowalla dataset.}
    \vspace{-0.15in}
    \begin{tabular}{ccccccc}
        \toprule
        Method & R@10 & R@20 & R@40 & N@10 & N@20 & N@40\\
        \midrule
        LightGCN  & 0.1526 & 0.2258 & 0.3199 & 0.1230 & 0.1451 & 0.1716\\
        SGL       & 0.1640 & 0.2369 & 0.3268 & 0.1321 & 0.1540 & 0.1795\\
        HCCF      & 0.1634 & 0.2370 & 0.3239 & 0.1305 & 0.1529 & 0.1778\\
        SimGCL    & 0.1605 & 0.2358 & 0.3335 & 0.1302 & 0.1529 & 0.1802\\
        NCL       & 0.1666 & 0.2441 & 0.3401 & 0.1342 & 0.1574 & 0.1844\\
        DirectAU  & 0.1533 & 0.2276 & 0.3221 & 0.1211 & 0.1439 & 0.1702\\
        LightGCL  & 0.1607 & 0.2387 & 0.3395 & 0.1285 & 0.1522 & 0.1807\\
        AutoCF    & 0.1747 & 0.2506 & 0.3482 & 0.1407 & 0.1635 & 0.1913\\
        DCCF      & 0.1672 & 0.2464 & 0.3476 & 0.1352 & 0.1589 & 0.1873\\
        GFormer   & 0.1730 & 0.2487 & 0.3456 & 0.1426 & 0.1651 & 0.1924\\    
        \bottomrule
    \end{tabular}
    \vspace{-0.05in}
    \label{tab:gcf_result}
\end{table}

\noindent \textbf{Experimental Settings.}
In the general collaborative filtering scenario, the objective of the recommender system is to suggest unexplored items to users by leveraging the collaborative insights gleaned from the existing user-item interaction data. In this study, we conduct experiments on the publicly available Gowalla dataset, which is obtained from the Gowalla platform and captures the check-in relationships between users and various locations.\\\vspace{-0.12in}

\noindent \textbf{Results and Analysis.} The experimental results of collaborative filtering methods on the Gowalla dataset are presented in Table~\ref{tab:gcf_result}. From these results, we can make the following key observations.
\begin{itemize}[leftmargin=*]

\item Graph neural networks (GNNs) are widely used in state-of-the-art collaborative filtering algorithms. GNNs are effective in capturing high-order collaborative filtering signals by propagating and aggregating information on the user-item interaction graph. This enables them to optimize recommendation performance by incorporating these signals into learned embeddings.

\item Methods such as SGL and AutoCF go beyond pure GNN-based algorithms like LightGCN and introduce contrastive or generative SSL objectives. These methods achieve significant performance improvement by leveraging additional supervision signals.

\item SGL constructs an enhanced contrastive perspective by randomly deleting graph structures. However, the randomness in this process can introduce unnecessary noise. To address this, update methods adopt feature-augmentation-based approaches (e.g., SimGCL, NCL) or automated self-supervised signals (e.g., AutoCF). These approaches effectively generate adaptive supervision signals, mitigating the impact of noise perturbation and leading to better recommendation performance.

\end{itemize}

\subsection{Sequential Recommendation}

\begin{table}
    \centering
    \small
    %\footnotesize
    %\scriptsize
    \caption{Performance Evaluation by \model\ for Sequential Recommender Systems on MovieLen-20M Dataset.}
    \vspace{-0.05in}
    \begin{tabular}{ccccccc}
        \toprule
        Method & R@5 & R@10 & R@20 & N@5 & N@10 & N@20\\
        \midrule
        BERT4Rec & 0.1493 & 0.2223 & 0.3197 & 0.1003 & 0.1237 & 0.1483\\
        CL4SRec  & 0.1491 & 0.2245 & 0.3221 & 0.1007 & 0.1250 & 0.1495\\
        DuoRec   & 0.1507 & 0.2242 & 0.3182 & 0.1021 & 0.1258 & 0.1494\\
        DCRec    & 0.1569 & 0.2327 & 0.3303 & 0.1067 & 0.1311 & 0.1557\\
        \bottomrule
    \end{tabular}
    \vspace{-0.05in}
    \label{tab:seqrec_result}
\end{table}

\noindent \textbf{Experimental Settings.} In the context of sequential recommendation, the objective of a recommender system is to predict the next item in users' interaction sequences by leveraging their historical behavior. This involves analyzing the user's past interactions to uncover their preferences and interests, which are then used to generate personalized recommendations. In our experiment, we assess the effectiveness of advanced sequential models on the MovieLens-20M dataset. The MovieLens dataset provides not only users' movie ratings but also the timestamps of their interactions. This valuable information allows us to construct sequences of movies that users have interacted with, enabling the recommender system to make predictions about the next item in the sequence.

Specifically, for each evaluated model (BERT4Rec, CL4SRec, DuoRec, and DCRec), we standardize the training process by setting the embedding dimension to 64 and the batch size to 512. We also ensure reproducibility of the training results by fixing the random seed. To optimize the model's performance, we utilize SSLRec's automated parameter search to determine the optimal combination of hyperparameters, such as the number of transformer layers and the dropout ratio. We train each model for a specific number of epochs on the training set, continuing until the model converges. After training, we evaluate each model's performance on the test set using Recall and NDCG as our evaluation metrics. \\\vspace{-0.12in}

\noindent \textbf{Results and Analysis.} The results depicted in Table~\ref{tab:seqrec_result} provide valuable insights into the current landscape of SSL-enhanced sequential recommenders. We summarize the following key observations.
\begin{itemize}[leftmargin=*]
\item One significant trend that emerges is the widespread adoption of the transformer architecture as the primary encoder in these methods. The transformer's multi-head self-attention mechanism plays a pivotal role in enabling seamless information transmission and integration for each item within a sequence. This empowers the models to discern user preferences and effectively encapsulate the holistic sequence information.

\item Among the compared algorithms, BERT4Rec stands out as a generative method that constructs self-supervised signals. It accomplishes this by randomly masking items in a sequence and reconstructing them. BERT4Rec serves as a fundamental baseline for self-supervised sequential recommendation. In contrast, CLS4Rec and DuoRec adopt distinct approaches to sequence augmentation, such as reordering and insertion, while also employing model-based enhancement techniques like multi-level dropout. These methods generate self-supervised signals through contrastive learning, resulting in improved performance. 

\item Notably, DCRec surpasses the other methods by not only leveraging sequence information for learning user preferences but also constructing an item transition/co-interaction graph. This graph allows the model to mine collaborative signals and generate contrastive learning views. Consequently, DCRec achieves the highest performance among the evaluated methods. In summary, the findings highlight the advancements made in leveraging transformer-based encoders, sequence augmentation techniques, and contrastive learning approaches to enhance the performance of sequence-based recommendation systems.

\end{itemize}

\subsection{Social Recommendation}

\begin{table}
    \centering
    \small
    %\footnotesize
    %\scriptsize
    \caption{Performance Evaluation by \model\ for Social Recommendation on Yelp (Social) Dataset.}
    \vspace{-0.05in}
    \begin{tabular}{ccccccc}
        \toprule
        Method & R@10 & R@20 & R@40 & N@10 & N@20 & N@40\\
        \midrule
        KCGN  & 0.0202 & 0.0355 & 0.0578 & 0.0094 & 0.0133 & 0.0178\\
        SMIN  & 0.0216 & 0.0368 & 0.0599 & 0.0103 & 0.0141 & 0.0188\\
        MHCN  & 0.0314 & 0.0525 & 0.0841 & 0.0163 & 0.0216 & 0.0280\\
        DSL   & 0.0335 & 0.0536 & 0.0846 & 0.0167 & 0.0218 & 0.0281\\
        \bottomrule
    \end{tabular}
    \vspace{-0.1in}
    \label{tab:socialrec_result}
\end{table}

\noindent \textbf{Experimental Settings.} Social Recommendation is a recommendation paradigm that exploits the social relationships between users, such as friendships, to enhance the accuracy of recommendations. The crux of this field lies in effectively mining valuable insights from user-user interactions and transforming the resulting social knowledge into meaningful representations. Ultimately, the efficacy of social recommendation hinges on the ability to fully leverage the informative power of social ties. To produce the leaderboard of this scenario, we train and evaluate the implemented social-based recommenders\footnote{When implementing the DcRec model on the Yelp dataset, we faced an Out-of-Memory (OOM) error despite using a powerful RTX 3090 GPU with 24 GB of memory, due to the high computational requirements.} (\ie, KCGN \cite{huang2021knowledge}, SMIN \cite{long2021social}, MHCN \cite{yu2021self}, and DSL \cite{wang2023denoised}) on the well-known public dataset Yelp.

Yelp dataset has user information about their friends, valuable for constructing a network for social recommendations. To ensure a fair comparison, we set the embedding dimension to 64 for all social methods. Each model is trained with a fixed number of epochs specific to its requirements. We evaluate the ranking-based performance measures, such as Recall and NDCG, for each method. Similar to other scenarios, we tune the hyper-parameters of each model using grid search to obtain the optimal performance. \\\vspace{-0.12in}

\noindent \textbf{Results and Analysis.} The results of the comparison of social-based recommenders are presented in Table~\ref{tab:socialrec_result}. Based on the evaluation results, we can draw the following observations.
\begin{itemize}[leftmargin=*]

\item Firstly, compared to the interaction information between users and items, user social relationships are known to contain a higher degree of noise \cite{yu2021self, wang2023denoised}. As a result, several algorithms have dedicated efforts to extract valuable structures from users' social connections while reducing the impact of noise on the recommender system's performance. By effectively filtering out irrelevant noise, these algorithms aim to enhance the quality of recommendations based on social relationships.

\item MHCN \cite{yu2021self} utilizes triangle motifs for hypergraph mining, enabling contrastive learning. KCGN \cite{huang2021knowledge} incorporates temporal information and heterogeneous user behavior to mitigate noise during feature learning. SMIN \cite{long2021social} employs meta-paths for secondary mining, identifying valuable neighboring nodes. DSL \cite{wang2023denoised} optimizes feature similarity between user pairs using Hinge Loss and utilizes self-supervised signals for social influence denoising.

\item Some techniques effectively combine global and local information in contrastive learning to reduce noise during mutual information maximization. For instance, MHCN incorporates sub-hypergraph representation to capture both overarching structure and local patterns in the social network. On the other hand, SMIN computes graph-level representation, combining global network information with local connections. The fusion of global and local information enhances model robustness, noise-handling capabilities, and improves recommendation accuracy and reliability.

\end{itemize}

\subsection{KG-enhanced Recommendation}

\begin{table}
    \centering
    \small
    %\footnotesize
    %\scriptsize
    \caption{Recall (R) and NDCG (N) reproduced by \model\ for KG-enhanced Recommendation on Alibaba dataset.}
    \vspace{-0.05in}
    \begin{tabular}{ccccccc}
        \toprule
        Method & R@5 & R@10 & R@20 & N@5 & N@10 & N@20\\
        \midrule
        KGIN  & 0.0746 & 0.1139 & 0.1678 & 0.0569 & 0.0711 & 0.0870\\
        KGCL  & 0.0749 & 0.1147 & 0.1685 & 0.0577 & 0.0720 & 0.0879\\
        KGRec & 0.0708 & 0.1188 & 0.1735 & 0.0595 & 0.0741 & 0.0903\\
        \bottomrule
    \end{tabular}
    \vspace{-0.15in}
    \label{tab:kgrec_result}
\end{table}

\noindent \textbf{Experimental Settings.} In the present study, we evaluate the performance of KG-enhanced recommendation algorithms, namely KGIN \cite{wang2021learning}, KGCL \cite{yang2022knowledge}, and KGRec \cite{yang2023knowledge}, using the Alibaba dataset. The objective is to leverage the structured information within a knowledge graph to enhance recommendation accuracy. To construct the knowledge graph, representative entities are sourced from the original data, and the approach suggested in \cite{tian2021joint} is employed to extract relevant information from Wikidata\footnote{\href{https://query.wikidata.org/}{https://query.wikidata.org/}}.

To assess the effectiveness of the KG-enhanced recommendation algorithms, we build a leaderboard by training and evaluating the ranking-based performance of each model on the MIND dataset and compare their performance against each other. For detailed hyperparameters, we standardize the embedding dimension to 64 for all KG-based models and train them with a fixed number of epochs specific to each algorithm. Additionally, we optimize the model-specific hyperparameters for each model using grid search. \\\vspace{-0.12in}

\noindent \textbf{Results and Analysis.} KG-enhanced recommendation leverages item-side knowledge graphs to enhance collaborative signals. The evaluation results are shown in Tabel~\ref{tab:kgrec_result}. 

\begin{itemize}[leftmargin=*]

\item The abundance of knowledge can introduce noise, which hinders its effective utilization. Mining item-specific information with rich content from the knowledge graph and transforming it into self-supervised signals can significantly enhance algorithmic performance in KG-enhanced recommendation.

\item Self-supervised KG recommenders, such as KGCL and KGRec, outperform KGIN by employing techniques like contrastive learning and rational masking. These approaches help identify significant item nodes, diminish the influence of noise, and explore the interrelation between knowledge graph and collaborative filtering signals for improved recommendations. \vspace{-0.15in}

\end{itemize}

\subsection{Multi-behavior Recommendation}

\begin{table}
    \centering
    \small
    %\footnotesize
    %\scriptsize
    \caption{Recall (R) and NDCG (N) reproduced by \model\ for Multi-behavior Recommendation on Retail Rocket dataset.}
    \vspace{-0.05in}
    \begin{tabular}{ccccccc}
        \toprule
        Method & R@10 & R@20 & R@40 & N@10 & N@20 & N@40\\
        \midrule
        MBGMN   & 0.0405 & 0.0478 & 0.0529 & 0.0227 & 0.0245 & 0.0255\\
        HMG-CR  & 0.0363 & 0.0446 & 0.0584 & 0.0163 & 0.0184 & 0.0211\\
        S-MBRec & 0.0336 & 0.0391 & 0.0593 & 0.0124 & 0.0138 & 0.0180\\
        CML     & 0.0428 & 0.0492 & 0.0570 & 0.0241 & 0.0257 & 0.0272\\
        KMCLR   & 0.0428 & 0.0501 & 0.0557 & 0.0240 & 0.0258 & 0.0269\\
        \bottomrule
    \end{tabular}
    \vspace{-0.15in}
    \label{tab:mbrec_result}
\end{table}

\begin{table*}[!tb]
    \centering
    \small
    %\footnotesize
    %\scriptsize
    \caption{Comparison with existing PyTorch-based recommender system frameworks.}
    \vspace{-0.1in}
    \resizebox{0.90\textwidth}{!}{
    \begin{threeparttable}
    \begin{tabular}{cccccccc}
        \toprule
        Framework & \#Models & \#SSL Models & \#Datasets & \#Scenarios & Release time & SSL-specific & Automated PT\\
        \midrule
        Spotlight~\cite{kula2017spotlight}   & 8  & 0  & 5  & 2  & 2017 & \color{red}{\XSolidBrush}  & \color{tgreen}{\Checkmark} \\
        DaisyRec~\cite{sun2020we}            & 13 & 0  & 14 & 1  & 2019 & \color{red}{\XSolidBrush}  & \color{tgreen}{\Checkmark} \\
        ReChorus~\cite{wang2020make}         & 18 & 2  & 4  & 2  & 2020 & \color{red}{\XSolidBrush}  & \color{red}{\XSolidBrush}  \\
        Beta-recsys~\cite{meng2020beta}      & 27 & 3  & 21 & 4  & 2020 & \color{red}{\XSolidBrush}  & \color{tgreen}{\Checkmark} \\
        RecBole~\cite{zhao2021recbole}       & 143& 11 & 41 & 9 & 2020 & \color{red}{\XSolidBrush}  & \color{tgreen}{\Checkmark} \\
        SELFRec~\cite{yu2022self}            & 16 & 11 & 4  & 2  & 2021 & \color{tgreen}{\Checkmark} & \color{red}{\XSolidBrush}  \\
        \midrule
        \textbf{SSLRec}      & \textbf{27} & \textbf{25} & \textbf{13} & \textbf{5}  & \textbf{2022} & \color{tgreen}{\Checkmark} & \color{tgreen}{\Checkmark} \\
        \bottomrule
    \end{tabular}
    \begin{tablenotes}    
   \footnotesize              
       \item[1] Automated PT denotes automated hyper-parameter tuning.  The statistics were collected on the date of Jul, 2023.
    \end{tablenotes}  
    \end{threeparttable}
    }
    \label{tab:comparison}
    \vspace{-0.15in}
\end{table*}

\noindent \textbf{Experimental Settings.} Multi-behavior recommender systems consider different user interactions with items to enhance recommendation accuracy. For example, in e-commerce, users perform various actions like viewing, adding to cart, and making purchases. The main goal is to optimize recommendations based on the most important behavior, which is purchasing. To evaluate the performance of multi-behavior recommendation, experiments were conducted using five different methods: MBGMN, HMG-CR, S-MBRec, CML, and KMCLR. These methods were tested on a real-world e-commerce dataset called Retail-rocket, which encompasses three distinct types of user behaviors. During the experiments, all models were meticulously trained on a dedicated training set until they reached convergence. The training process involved careful exploration and selection of optimal hyperparameters. Subsequently, the trained models were evaluated on a separate test set using ranking-based metrics to assess their performance in recommendations. \\\vspace{-0.12in}

\noindent \textbf{Results and Analysis.} Table~\ref{tab:mbrec_result} presents the performance of multi-behavior recommendation algorithms on the Retail-rocket dataset. The effectiveness of these algorithms lies in uncovering intricate patterns in users' complex behaviors. \vspace{-0.05in}

\begin{itemize}[leftmargin=*]

\item MBGMN employs a meta-learning approach to model users' multi-behavior information. However, due to the sparsity of observed user behavior label data, it falls short in achieving optimal modeling outcomes and performs inferiorly compared to self-supervised methods like CML and KMCLR.

\item HMG-CR, S-MBRec, CML, and KMCLR methods use contrastive learning to improve the prediction of the target behavior by increasing mutual information between the target behavior and auxiliary behaviors. HMG-CR constructs a hyper meta-graph, while S-MBRec, CML, and KMCLR extract useful aspects from different behaviors to enhance performance. These techniques enhance recommendations compared to MBGMN by leveraging multiple user behaviors to optimize recommendations.

\item KMCLR incorporates a knowledge graph and applies knowledge-aware contrastive loss to learn robust item features, mitigating the impact of data sparsity and improving the effectiveness of multi-behavior recommendation. These advanced techniques enable KMCLR and other methods to achieve more precise and effective recommendations, even in scenarios with sparse data. \vspace{-0.1in}

\end{itemize}

%% file: relatedwork.tex
\section{Comparison with existing frameworks}
\label{sec:relatedwork}

Deep neural learning has transformed recommender systems, leading to the development of numerous recommender frameworks~\cite{bin2019neurec, kula2017spotlight, sun2020we, wang2020make, meng2020beta, zhao2021recbole, yu2021self}. This section focuses on comparing and discussing PyTorch-based recommender frameworks known for their ease of use and implementation of the latest algorithms. Table~\ref{tab:comparison} provides statistics on these frameworks, revealing key observations.

\begin{itemize}[leftmargin=*]

\item \textbf{Rich SSL Functions and Modules.} 
Despite the impressive performance of self-supervised recommendation algorithms in various user modeling and personalization scenarios and their prevalence in recent publications~\cite{yu2022self}, most existing recommendation frameworks do not comprehensively address this topic. Early frameworks such as Spotlight~\cite{kula2017spotlight} and DaisyRec~\cite{sun2020we} mainly focused on classical supervised recommendation algorithms, while ReChorus~\cite{wang2020make}, Beta-recsys~\cite{meng2020beta}, and RecBole~\cite{zhao2021recbole} implement a wide range of algorithms, including some state-of-the-art self-supervised recommendation algorithms. However, it was not until the recent emergence of SELFRec~\cite{yu2022self} that a framework specifically designed for self-supervised learning is introduced. Our framework addresses an overlooked topic in recommendation frameworks and provides customized functions and modules for SSL. This offers users easy-to-use functions and flexibility. With a rich set of SSL functions and modules, our framework enables quick implementation of various SSL recommendation algorithms, benefiting researchers and practitioners. \\\vspace{-0.12in}

\item \textbf{Diverse Recommendation Scenarios and SOTA Models.} When comparing to existing algorithm frameworks, \model\ stands out as it provides the most comprehensive collection of self-supervised recommendation algorithms. The majority of these algorithms have achieved state-of-the-art performance in the field. While SELFRec, another self-supervised framework, primarily focuses on general collaborative filtering and sequential recommendation, our \model\ goes beyond by covering five distinct tasks and implementing a larger number of self-supervised recommendation algorithms. Additionally, \model\ provides auxiliary toolkits like automated hyper-parameter search, result saving, and process logging, which enhance usability and reliability. \\\vspace{-0.15in}

\end{itemize}

%% file: conclusion.tex
\section{Conclusion}
\label{sec:conclusoin}

In this paper, we introduce \model, a comprehensive framework for self-supervised recommendation algorithms. Through modular algorithm modeling, we facilitate rapid design and development. Our framework includes commonly used augmentation and self-supervised loss modules, simplifying the algorithm design process. We offer standardized datasets and consistent training-validation-testing procedures for fair comparisons. We value user feedback and strive to enhance and expand \model\ to meet the evolving needs of the recommendation system research community. We will improve and optimize the \model\ framework based on valuable user feedback, with a focus on self-supervised recommendation.